\documentclass [12pt]{article}

\begin{document}

\begin{center}
{\large\bf SPLITTING OF RESONANT FREQUENCIES OF ACOUSTIC WAVES IN ROTATING COMPRESSIBLE FLUID}\\
\medskip
 A. N. Tarasenko, A. A. Sokolsky\\
\textit{Belarusian State University, 4 Nezalejnosty Ave., 220030 Minsk, Belarus,}\\
 \textit{E-mail: sokolan@tut.by}
\end{center}

\medskip
\begin{center}
\begin{minipage}{0.8\textwidth}
{\footnotesize It is shown that in a rotating compressible fluid the resonant frequencies (measured
in a system of reference rotating together with the medium) for the azimuthally running acoustic
waves are split into two components. The received results can be of practical significance as a
basis of a method of measurements of angular speed of medium and for acoustics of rotating
technical devices.}
\end{minipage}
\end{center}

As is known [1], in a compressible fluid, localized between two rigid
coaxial cylindrical walls, free oscillations of the character of both
standing and running in an azimuthal direction waves are possible. With the
help of the equations

\begin{equation}
\label{eq1}
\rho _0 \dot {\vec {V}} = - \vec{\nabla} \tilde {p}\,, \quad \dot {\tilde {p}} = - \rho _0 c_0^2
\,\mathop{div}\vec {V}
\end{equation}

\noindent
it is easy to show that for the standing waves, in the absence of axial
component of oscillation speed $\vec {V}$ (i. e., for $V_z = 0)$, the
acoustic pressure $\tilde {p} = p - p_0 $ looks like $\tilde {p}(r,\varphi
,t) = P_m (r)\cos m\,\varphi \;\cos \omega _{mn} t$, and for the azimuthally
running waves $\tilde {p}(r,\varphi ,t) = P_m (r)\cos (m\,\varphi - \omega
_{mn} t)$. Here, $P_m (r) = A\;J_m (\omega _{mn} r / c_0 ) + B\;Y_m (\omega
_{mn} r / c_0 )$, $ A$ and $B$ are the constants, $r,\varphi ,\;z$ are the
cylindrical coordinates, $J_m $ and $Y_m $ are the Bessel functions [2], $m$ è
$n$ are the integers, $p_0 $, $\rho _0 $ and $c_0 $ are the equilibrium
pressure, density and sound speed. The sets of resonant frequencies $\omega
_{mn} $ for both cases are identical and can be found from the equation

\begin{equation}
\label{eq2}
{J}'_m \left( {\omega R_1 / c_0 } \right){Y}'_m \left( {\omega R_2 / c_0 }
\right) = {J}'_m \left( {\omega R_2 / c_0 } \right){Y}'_m \left( {\omega R_1
/ c_0 } \right),
\end{equation}

\noindent
where $R_{1}$ and $R_{2}$ are the radiuses of the walls, ${J}'_m \left( x
\right) = dJ_m / dx,\;\;{Y}'_m \left( x \right) = dY_m / dx$.

Let us consider the behaviour of such waves on the background of the
equilibrium state of the medium rotating with a constant angular speed $\vec
{\Omega }$. Nondissipative hydrodynamic processes in the rotating compressed
liquid (gas), considered from the point of view of the system of reference S
rotating together with the fluid, are described by system of equations

\begin{equation}
\label{eq3}
\begin{array}{l}
 \dot {\vec {V}} + (\vec {V} \cdot \vec {\nabla })\vec {V} = - \,\rho ^{ -
1}\vec{\nabla}p + \Omega ^2\,\vec {r} + 2\,[\vec
{V}\times \vec {\Omega }], \\
 \dot {\rho } + \mathop{div}(\rho \vec {V}) = 0,\mbox{ }\;\;\,\quad \dot {p} + \vec
{V} \cdot \vec {\nabla }p = c^2(\dot {\rho }\, + \vec {V} \cdot \vec {\nabla
}\rho ), \\
 \end{array}
\end{equation}

\noindent where $\vec {V}$ is the velocity of a medium element relative to S. Within the limits of
linear acoustics and for relatively low speeds of rotation (i. e., for small value of $M = \Omega R
/ c)$, system (\ref{eq3}), in view of a condition of equilibrium $\vec{\nabla}p_0 = \rho _0 \Omega
^2\,\vec {r}$, leads (instead of (\ref{eq1})) to the following equations:

\begin{equation}
\label{eq4}
\dot {\vec {V}} = - \rho _0^{ - 1} \vec{\nabla} \tilde {p} + 2[\vec {V}\times \vec {\Omega }],
\end{equation}

\begin{equation}
\label{eq5}
\dot {\tilde {p}} = - \rho _0 c_0^2 \,\mathop{div}\vec {V}.
\end{equation}

System (\ref{eq4}), (\ref{eq5}) describes adequately the corrections of the first order in
$M$, caused by the influence of the medium rotation on oscillations, if the
conditions $M \ll 1$ and $V_A / c \sim M^2$ are satisfied ($V_A $ is the
amplitude of the velocity of oscillations). Thus the ratio $\Omega / \omega
$ for the resonant values of $\omega $ will be of the order $M$, and $p_0 $,
$\rho _0 $, and $c_0 $ in equations (\ref{eq4}), (\ref{eq5}) can be considered constant.

This allows to use system (\ref{eq4}), (\ref{eq5}), for example, for the frequencies of
rotation of the order 1 -- 10 r.p.s. and the radial sizes 0.2 -- 0.5 m for
the sound intensity corresponding to the values $V_A / c < 10^{ - 5}$ for
both gases and liquids.

It is easy to check that, in the approximation considered, elimination of
$\vec {V}$ from (\ref{eq4}), (\ref{eq5}) leads to the same differential equation for the
acoustic pressure $\tilde {p}$ as in the case of non-rotating medium:

\begin{equation}
\label{eq6}
\ddot {\tilde {p}} = c_0^2 \;\nabla ^2\,\tilde {p},
\quad
\nabla ^2\, = \frac{1}{r}\frac{\partial }{\partial r} + \frac{\partial
^2}{\partial r^2} + \frac{1}{r^2}\frac{\partial }{\partial \varphi ^2} +
\frac{\partial ^2}{\partial z^2}_{.}
\end{equation}

At the same time, the boundary condition for $\tilde {p}(r,\varphi ,t)$ on a
rigid cylindrical wall of the radius $R$ (that follows from the requirement
$V_r (r = R) = 0)$ gets, by virtue of (\ref{eq4}), an additional term, which have
the first order in $\Omega / \omega $:

\begin{equation}
\label{eq7}
\left( {\frac{\partial \dot {\tilde {p}}}{\partial r} + \frac{2\Omega
}{r}\frac{\partial \tilde {p}}{\partial \varphi }} \right)_{r = R} = 0.
\end{equation}

As for the azimuthally running wave (i. e., for the harmonic of number $m =
\pm 1,\;\pm 2,\;\pm 3$\ldots ) the acoustic pressure $\tilde {p}(r,\varphi
,t)$ looks like $P_m (r)\cos \,(m\,\varphi - \omega t)$, it follows from
condition (\ref{eq7}):

\begin{equation}
\label{eq8}
\left( {\frac{dP_m }{dr} - \frac{2\Omega m}{\omega r}P_m } \right)_{r =
R\;_{1,\;2} } = 0,
\end{equation}

\noindent
where $R_{1}$ and $R_{2}$ are the radiuses of rigid coaxial cylindrical walls
between which the wave is localized.

According to equation (\ref{eq6}), when $\Omega \ne 0$, $P_m (r)$ holds the form
$A\;J_m (\omega r / c_0 ) + B\;Y_m (\omega r / c_0 )$. Substituting $P_m
(r)$ in (\ref{eq8}), we receive, instead of (\ref{eq2}), the condition

\begin{equation}
\label{eq9}
\frac{\displaystyle {J}'_m \left( {\frac{\omega R_1}{c_0}} \right) - \frac{2mc_0\Omega }{\omega
^2R_1 }J_m
\left( {\frac{\omega R_1}{c_0}} \right)}{\displaystyle {J}'_m \left( {\frac{\omega R_2}{c_0}}
\right) - \frac{2mc_0\Omega }{\omega ^2R_2 }J_m \left( {\frac{\omega R_2}{c_0}}
\right)} = \frac{\displaystyle {Y}'_m \left( {\frac{\omega R_1}{c_0}} \right) - \frac{2mc_0\Omega
}{\omega ^2R_1 }Y_m \left( {\frac{\omega R_1}{c_0}} \right)}{\displaystyle {Y}'_m \left(
{\frac{\omega R_2}{c_0}} \right) - \frac{2mc_0\Omega }{\omega ^2R_2 }Y_m \left( {\frac{\omega
R_2}{c_0}}
\right)}.
\end{equation}

Values of $\omega $ satisfying (\ref{eq9}) form a spectrum of the resonant
frequencies of the azimuthally running waves: $\omega _{mn} $ at $\Omega =
0$, $\omega _{mn}^ + $, if $m\,\Omega > 0$ (i. e., for a wave running in a
direction of medium rotation) and $\omega _{mn}^ - $, if $m\,\Omega < 0$.

Frequencies $\nu _{mn} = \,2\pi \omega _{mn} $ and frequency deviations $\nu _{mn}^ + - \nu _{mn} $
and $\nu _{mn}^ - - \nu _{mn} $, where $\nu _{mn}^\pm = 2\pi \omega _{mn}^\pm $ obtained by a
numerical solution of equation (\ref{eq9}) at $_{ }R_{1}$ = 0.2 m, $_{ }R_{2}$ = 0.5 m, $\Omega =
\pm 2\pi N$, $N$ = 10 r.p.s, and $c_{0}$ = 330 m/s (for $m $= 1, 2, 3; $n $=1, 2, 3, 4, 5) are
presented in Table 1.

\begin{table}[htbp]
\begin{tabular}
{|c|p{56pt}|p{56pt}|p{56pt}|p{56pt}|p{56pt}|} \hline m$\backslash $n& \hfil 1 \hfil&\hfil 2 \hfil&
\hfil 3 \hfil&\hfil 4 \hfil&\hfil 5 \hfil
 \\
\hline
 1&
 \hfil 153.55 \hfil \par  -3.91 \hfill 1.32&
 \hfil 594.45 \hfil \par 1.41 \hfill - 1.44&
 \hfil 1122.19 \hfil \par 0.43 \hfill - 0.43&
 \hfil 1664.72 \hfil \par 0.20 \hfill - 0.20&
 \hfil 2211.01 \hfil \par 0.11 \hfill - 0.11 \\
\hline 2&
 \hfil 298.57 \hfil \par - 5.24 \hfill 4.06&
 \hfil 673.95 \hfil \par 2.29 \hfill - 2.39&
 \hfil 1161.34 \hfil \par 0.86 \hfill - 0.87&
 \hfil 1690.30 \hfil \par 0.40 \hfill - 0.40&
 \hfil 2230.04 \hfil \par 0.23 \hfill - 0.23 \\
\hline 3&
 \hfil 431.53 \hfil \par  -6.12 \hfill  5.43&
 \hfil 791.52 \hfil \par 2.25 \hfill - 2.40&
 \hfil 1225.43 \hfil \par 1.27 \hfill - 1.29&
 \hfil 1732.52 \hfil \par 0.60 \hfill - 0.61&
 \hfil 2261.55 \hfil \par 0.34 \hfill - 0.34 \\
\hline
\end{tabular}
\label{tab1}
\caption[]{Frequencies $\nu _{mn} $(Hz) and deviations: $\nu _{mn}^ + - \nu _{mn} $, $\nu _{mn}^ - -
\nu _{mn} $ (Hz).}
\end{table}

For a cylinder of the radius $R$ (i. e., for $R_{1 }$= 0, $R_{2 }=R)$ the function $P_m (r)$ is
reduced to $A\;J_m (\omega \,r / c_0)$, and we can receive the approximated expression for the
resonant frequencies in an explicit form:

\begin{equation}
\label{eq10}
\omega _{mn} + \Delta \omega _{mn} = \frac{c_0}{R}X_{mn} + \frac{2\,m\;\Omega \;J_m \left( {X_{mn}
}
\right)}{X_{mn}^2 \;{J}''_m \left( {X_{mn} }
\right)},
\end{equation}

\noindent
where $X_{mn} $ are the roots of the equation $\;{J}'_m (x) = 0$.

According to (\ref{eq10}), ratio of splitting of frequencies to angular speed of
medium in this case is of a universal character:

\begin{equation}
\label{eq11}
f_{mn} = \frac{\omega _{mn}^ + - \omega _{mn}^ - }{\Omega } =
\frac{4\,m\;J_m \left( {X_{mn} } \right)}{X_{mn}^2 \;{J}''_m \left( {X_{mn}
} \right)}.
\end{equation}

Table 2 gives the values of the roots $X_{mn} $ and of the ``splitting
coefficients'' $f_{mn} $.

\begin{table}[htbp]
\begin{tabular}
{|c|p{60pt}|p{60pt}|p{60pt}|p{60pt}|p{60pt}|}
 \hline m$\backslash $n& \hfil1\hfil&\hfil 2\hfil& \hfil3\hfil&\hfil 4\hfil& \hfil5 \\
\hline 1& 1.841 \par - 1.674& 5.331 \par - 0.146& 8.536 \par - 0.056& 11.706 \par - 0.029&
14.864 \par - 0.018 \\
\hline 2& 3.054 \par - 1.501& 6.706 \par - 0.195& 9.969 \par - 0.084& 13.170 \par - 0.047&
16.348 \par - 0.030 \\
\hline 3& 4.201 \par - 1.387& 8.015 \par - 0.217& 11.346 \par - 0.100& 14.586 \par - 0.059&
17.789 \par - 0.039 \\
\hline
\end{tabular}
\label{tab2}
\caption[]{Values of $X_{mn} $ and $f_{mn} $ for $ m$ = 1, 2, 3; $ n$ = 1, 2, 3, 4, 5.}
\end{table}

As the frequencies of the running waves following and opposing a direction
of rotation are not equal, it is impossible to construct a standing wave as
some their superposition. The absence of standing waves for $m\;\Omega \ne
0$ follows directly from impossibility to satisfy boundary conditions (\ref{eq7}) by
the pressure $\tilde {p}(r,\varphi ,t)$ of the form $P_m (r)\cos m\,\varphi
\;\cos \omega t$. For $m\;\Omega \ne 0$, a pulse of waves moving in an
azimuthal direction with the angular speed $\Omega _{mn}^{puls} $, which is
equal to $(\omega _{mn}^ + - \omega _{mn}^ - ) / (2m)$ relative to rotating
medium, will occur instead of a standing wave.

The splitting of sound frequencies in a rotating medium can be, in some
sense, considered as an acoustic analog of the Sagnac effect in optics [3].

For values of parameters $M = \Omega R / c$ and $V_A / c$, which are rather
greater than considered above, taking into account of nonlinear members in
equations and corresponding correction of the received results are
necessary. We expect, however, that, at least for weak nonlinearity, the
effect of splitting of frequencies will be of similar character, and that
the resonant response of medium will begin to be also revealed on the
frequencies of pulses.


\begin{thebibliography}{99}
\bibitem{l1} Isakovich M. A. General Acoustic. Moscow. 1973. P. 271.
\bibitem{l2} Handbook of mathematical functions. Edited by M. Abramovitz, I. Stegun, NBSAMS, 1964.
\bibitem{l3} Landau L. D., Lifshits E. M. Theory of Field. Moscow. 1988. P. 326.
\end{thebibliography}
\end{document}